\documentclass[useAMS,usenatbib]{mn2e}
\usepackage{graphicx}
\usepackage{times}
\usepackage{amssymb}
\title[Dust in The Wind]{Intergalactic Dust Extinction in Hydrodynamic 
Cosmological Simulations}
\author[Zu et al.]{ 
\parbox{\textwidth}{
Ying Zu$^1$\thanks{E-mail: yingzu@astronomy.ohio-state.edu},
David H. Weinberg$^{1,2,3}$, 
Romeel Dav\'{e}$^{4}$, 
Mark Fardal$^{5}$,
Neal Katz$^{5}$, 
Du{\v s}an Kere{\v s}$^{6}$,
Benjamin D. Oppenheimer$^{7}$
}
\vspace*{4pt} \\
${1}$Department of Astronomy, The Ohio State University, 140 W. 18th Avenue,
Columbus, OH 43210, USA\\
${2}$Center for Cosmology and Astro-Particle Physics, The Ohio State
University, Columbus, OH 43210\\
${3}$Institute for Advanced Study, Princeton, NJ 08540\\
${4}$Astronomy Department, University of Arizona, Tucson, AZ 85721, USA\\
${5}$Department of Physics and Astronomy, University of
Massachusetts at Amherst, Amherst, MA
98105\\
${6}$Institute for Theory and Computation, Harvard-Smithsonian
Center for Astrophysics, Cambridge, MA 02138, USA\\
${7}$Leiden Observatory, Leiden University, 2300 RA Leiden,
The Netherlands
}

\def\extbar{\langle_{\rm }E(g-i)_{\rm }\rangle_g}

\def\md{\Sigma_d(r)}
\def\wp{w(r_p)}

\def\Msol{M_\odot}
\def\hMpc{h^{-1}{\rm Mpc}}
\def\hKpc{h^{-1}{\rm kpc}}

\def\perhMpcc{h^3{\rm Mpc}^{-3}}
\def\la{\langle}

\def\K{\,{\rm K}}
\def\cm{\,{\rm cm}}
\newcommand{\aap}{A{\&}A}

\begin{document} \date{\today} \maketitle
\begin{abstract} Recently M\'enard et al.\ (hereafter MSFR) detected a subtle
    but systematic change in the mean color of quasars as a function of their
    projected separation from foreground galaxies, extending to comoving
    separations of $\sim 10\hMpc$, which they interpret as a signature of
    reddening by intergalactic dust. We present theoretical models of this
    remarkable observation, using smoothed particle hydrodynamic (SPH)
    cosmological simulations of a $(50\hMpc)^3$ volume.  Our primary model
    uses a simulation with galactic winds and assumes that dust traces the
    intergalactic metals. The predicted galaxy-dust correlation function is
    similar in form to the galaxy-mass correlation function, and reproducing
    the MSFR data requires a dust-to-metal mass ratio of 0.24, about half the
    value in the Galactic ISM. Roughly half of the reddening arises in dust
    that is more than $100\hKpc$ from the nearest massive galaxy. We also
    examine a simulation with no galactic winds, which predicts a much
    smaller fraction of intergalactic metals (3\% vs.\ 35\%) and therefore
    requires an unphysical dust-to-metal ratio of 2.18 to reproduce the MSFR
    data.  In both models, the signal is dominated by sightlines with
    $E(g-i)=0.001-0.1$.  The no-wind simulation can be reconciled with the
    data if we also allow reddening to arise in galaxies up to several
    $\times 10^{10} M_\odot$.  The wind model predicts a mean visual
    extinction of $\langle A_V \rangle \approx 0.0133$ mag out to $z=0.5$,
    with a sightline-to-sightline dispersion similar to the mean, which could
    be significant for future supernova cosmology studies.  Reproducing the
    MSFR results in these simulations requires that a large fraction of ISM
    dust survive its expulsion from galaxies and its residence in the
    intergalactic medium.  Future observational studies that provide higher
    precision and measure the dependence on galaxy type and environment will
    allow detailed tests for models of enriched galactic outflows and the
    survival of intergalactic dust.  \end{abstract}
%
\section{Introduction} \label{sec:intro}
Heavy element enrichment in intergalactic gas has long been studied via
absorption lines in quasar spectra and, in more restricted regimes, via the
influence of metal species on the X-ray emission spectra of clusters and
groups.  In the Galactic interstellar medium (ISM), roughly half of the mass
of heavy elements is in solid phase --- interstellar dust --- which is thought
to contain roughly 2/3 of interstellar carbon and the great majority of
interstellar magnesium, silicon, and iron (\citealt{Draine09} and references
therein).  This solid-phase material is much more difficult to study in the
intergalactic medium (IGM), in part because the expected level of dust
extinction along any given line of sight is small, and in part because the
continuum spectra of potential background sources --- quasars, galaxies, and
supernovae --- show much greater variation than those of spectroscopically
typed stars.  In a recent breakthrough study, M{\'e}nard et al.\ (2009,
hereafter MSFR) used large samples of quasars and galaxies from the Sloan
Digital Sky Survey (SDSS; \citealt{York00}) to measure reddening by
intergalactic dust, detecting a systematic change in quasar colors with
increasing distance from foreground galaxies, measured to comoving projected
separations of $\sim 10\hMpc$.  In this paper, we use hydrodynamic
cosmological simulations to provide the first theoretical models of the MSFR
results and to infer their implications for galactic outflows and the survival
of dust in the IGM.

Ever since \cite{Chandrasekhar52} modeled the opacity of interstellar matter
using angular correlation statistics in the brightness fluctuations of the
Milky Way, there have been efforts to constrain the presence of a diffuse dust
component outside the Galaxy by measuring the extinction of background
objects.  \cite{Zwicky62} first claimed the existence of intracluster dust in
Coma based on the extinction of light from background clusters, though his
inferred level is certainly incompatible with more modern constraints.
Quasars are powerful ``backlights'' for studying the transparency of the
Universe across a wide range of cosmic time.  \cite{OH84} modeled the effects
of uniform and clumped dust distributions on the redshift evolution of quasar
luminosity functions, but they concluded that observations did not clearly
favor one specific dust distribution over another.  Taking advantage of the
efficiency and accurate calibration afforded by CCD observations,
\cite{Zaritsky94} averaged hundreds of background galaxies to detect a mean
color excess of $E(B-I)\sim0.067$ behind two nearby galactic halos, arguing
for the existence of dust out to at least $60\hKpc$. \cite{Menard08} also
found evidence of dust in the halos of L$_*$ galaxies via studies of MgII
absorbers. Confirming both findings, \cite{Kaneda09} presented a direct
detection of extended far-IR dust emission in the halo of NGC~253. The
existence of dust in the intra{\it cluster} medium has remained controversial,
however, as reviewed by \cite{Muller08}, who derived $E(B-R) = 0.005 \pm
0.008$ from galaxies behind 458 RCS clusters.  Using a larger sample of $\sim
10^4$ clusters from the SDSS, \cite{Chelouche07} obtained significant
detections of $E(g-i) \approx\,{\rm few}\times 10^{-3}$ out to several virial
radii using the average colors of background quasars.  \cite{Bovy08} inferred
upper limits at roughly the same level for a sample of $z\approx 0.05$ SDSS
clusters using the spectra of early-type galaxies, though their results are
not clearly inconsistent with Chelouche et al.'s.

MSFR instead examined the mean colors of 85,000 photometrically identified
quasars \citep{Richards04} as a function of angular separation from 20 million
foreground galaxies in a 3800 square degree sky area from the SDSS.  They also
used the quasar brightnesses to infer the galaxy-mass correlation function
from weak lensing magnification.  They found that the galaxy-dust and
galaxy-mass correlation functions track each other remarkably well, with a
correlated dust surface density profile $\Sigma_{\rm dust} \propto
\theta^{-0.8}$ detected out to angular separations that correspond to $\sim
10\hMpc$.  The color dependence of the reddening signal is consistent with
``standard'' interstellar dust, though the constraints are loose, implying
$R_V \equiv A_V/E(B-V) = 3.9 \pm 2.6$.  The MSFR results provide direct
evidence for extended intergalactic dust absorption, with an inferred total
dust mass comparable to the amount of dust in galaxy disks.  

Our primary model in this paper is based on a cosmological smoothed particle
hydrodynamics (SPH) simulation that incorporates the momentum-driven galactic
wind prescription of \cite{OD06}, which successfully matches several observed
aspects of gas-phase metal enrichment in galaxies and the IGM (see references
in \S\ref{sec:sim}).  In momentum-driven wind models, radiation pressure on
dust grains is the primary mechanism that sweeps gas out of galaxies
\citep{MQT05}.  We do not attempt to model the survival of dust in the IGM;
rather, we treat the ratio of dust mass to heavy element mass as a free
parameter to be inferred by matching the MSFR reddening measurements.  The
metallicity of the outflowing gas is assumed to equal that of the star-forming
gas, whose associated star formation effectively powers the outflow with
radiation pressure.  The model here is almost the converse of the one treated
by \cite{Aguirre01}, who considered radiative expulsion of dust (without gas
entrainment) and subsequent destruction as a source of IGM {\it gas}
enrichment.  As alternatives to the momentum-driven wind scenario, we also
consider two models based on an SPH simulation without galactic winds.  The
first assumes that the MSFR reddening signal arises in the simulation's
enriched intergalactic gas, which comes mainly from ram pressure and tidal
stripping in groups and clusters.  The second model allows additional
absorption in low mass galaxies.

We hope to address a number of questions inspired by the MSFR observations.
Can the simulation with galactic winds explain these observations?  What
dust-to-metal mass ratio is required to do so, and what does this imply about
the survival of dust in the IGM?  How is the dust distributed in the
simulation?  Do the models without galactic winds provide a viable alternative
explanation of the MSFR findings?  What further observations can test the model
predictions and provide greater insight into the origin and evolution of
intergalactic dust?

Section~\ref{sec:sim} describes our simulations and presents visual maps of the
predicted spatial distribution of intergalactic metals.  Section~\ref{sec:res}
presents our main results, including the predicted galaxy-dust correlations and
projected reddening profiles, the contributions to the reddening signal from
gas at different distances from high mass galaxies, the relative amounts of
galactic and intergalactic dust, and the cumulative distribution functions of
dust reddening. In section~\ref{sec:dis}, we briefly discuss the issues of
dust survival and extinction of high-redshift supernovae.
Section~\ref{sec:conc} summarizes our results with a look to potential future
directions for observational studies.

\section{Simulations and Metal Distributions} \label{sec:sim}
The simulations are performed using our modified \citep{OD08} version of
\rm{GADGET-2}~\citep{Springel05}, which combines a tree-particle-mesh
algorithm for gravitational calculations with smoothed particle hydrodynamics
(SPH, \citealt{Lucy77,GM77}). The cosmological parameters are $\Omega_{\rm
m}=0.25, \Omega_{\Lambda}=0.75, \Omega_{\rm b}=0.044,
H_0=70\,$km~s$^{-1}$Mpc$^{-1}, \sigma_8=0.8$ and $n=0.95$, in good agreement
with 5-year {\em Wilkinson Microwave Anisotropy Probe} (WMAP)
results~\citep{Hinshaw09}. We simulated the evolution of $288^3$ dark matter
particles and the same number of gas particles in a periodic box with a
comoving size 50$\hMpc$ on a side. The gas particle mass is $m_{\rm
SPH}=9\times10^7\Msol$, and the gravitational spline force softening is
$\epsilon = 4.9\hKpc$~(comoving; equivalent to Plummer softening $\epsilon =
3.5\hKpc$). We identify galaxies using the program {\rm
SKID}\footnote{http://www-hpcc.astro.washington.edu/tools/skid.html}, which
selects groups of stars and cold~($T<3\times10^4\K$),
dense~($\rho/\bar{\rho}_{\rm baryon}>1000$) gas particles that are associated
with a common density maximum.  Convergence tests indicate that the
simulations resolve galaxies with baryonic mass (stars plus cold gas) greater
than $\sim 64m_{\rm SPH}$, corresponding to $5.8\times10^9\Msol$.

One of our simulations is similar to the L50/288 run analyzed by Kere{\v s} et
al.\ (2009a,b), with slightly different cosmology. Star formation happens in a
sub-resolution two-phase medium, where thermal energy deposited by the
supernovae pressurizes the gas but does not drive outflows. We refer to the
intergalactic dust distribution of this simulation as the ``No-Wind Model.''

Our second simulation uses the same cosmological and numerical parameters as
the first one, but it incorporates the kinetic feedback wind mechanism of
GADGET-2 with the ``momentum-driven wind'' scalings of~\cite{OD06}; we refer to
the intergalactic dust distribution of this simulation as the ``Wind Model.''
Motivated by the analytical model of~\cite{MQT05}, this second simulation
scales the wind velocity with the velocity dispersion $\sigma$ of the galactic
halo and the mass loading factor (the ratio of gas ejection rate to star
formation rate) with $\sigma^{-1}$. The implementation is discussed in more
detail by~\cite{OD08}, though the simulation in that paper has slightly
different parameters.  Simulations with this wind implementation successfully
match observations of early IGM enrichment~\citep{OD06}, the galaxy
mass-metallicity relation~\citep{FD08}, OVI absorption at low
redshift~\citep{OD09}, enrichment and entropy levels in galaxy
groups~\citep{Dave08}, and the sub-L$_*$ regime of the galaxy baryonic mass
function~\citep{Oppenheimer10}.  However, there are both physical and numerical
uncertainties in this wind implementation, so we take it as a representative
illustration of how galactic winds could influence intergalactic dust. For the
purposes of our investigation, the key features of the wind model are the total
amount of metals expelled and the typical scale over which they are
distributed.
\begin{table} 
\caption[]{Gas-phase Metal Mass in the Two Simulations}
\centering
\begin{tabular}{lcc} 
\hline 
Model & All $Z$~($10^{10}\Msol$) & free $Z$~($10^{10}\Msol$)\\
\hline
Wind    & 288.0 & 100.7~(35.0\%) \\ 
No-Wind & 277.9 & 8.2~(3\%)      \\
\hline 
\end{tabular}
\label{tab:tab1}
\end{table}

The median redshift of the MSFR galaxy sample is $\langle_{\rm }z_{\rm
}\rangle\simeq0.36$, and the effective luminosity\footnote{The effective
luminosity is defined as the metallicity weighted mean luminosity of the
galaxy sample, assuming the metallicity-luminosity relation of
\cite{Tremonti04} and a Schechter luminosity function with $\alpha=-1.1$.} is
$\sim 0.45L_*$, with a corresponding comoving space density of $n_g\sim
0.01\perhMpcc$~\citep{Blanton03}.  To reasonably approximate the MSFR sample,
we analyze the $z=0.3$ redshift outputs of the two simulations and apply a
galaxy baryonic mass cut of $5.4\times10^{10} \Msol$~($\sim 600 m_{\rm SPH}$),
which yields $n_g = 0.01\perhMpcc$ in the Wind Model. We apply the same galaxy
baryonic mass cut to the No-Wind Model, though in this case the comoving space
density is higher, $n_g = 0.033\perhMpcc$. We enforce the same mass cut rather
than the same number density for the two galaxy samples so that we are
comparing two similar populations of galaxies~(with typical luminosity $\sim
0.45L_*$) in different simulations.\footnote{If we instead select a sample
with $n_g \sim 0.01\perhMpcc$ in the No-Wind simulation, the galaxy-mass
correlation rises by 1.4; our main conclusions about the No-Wind model are
immune to rescaling the galaxy-dust correlation function by such a factor.}
Note that we quote comoving distances throughout the paper, and unless
otherwise stated, the ``galaxies'' we refer to are those above the mass
threshold.  Since we are interested in modeling the reddening signal that is
{\it correlated} with galaxies, we do not need to include extinction that
might arise at lower or higher redshift than our simulation box, which would
produce a mean shift and dispersion in quasar colors that is uncorrelated with
the galaxy population in our simulation volume.

\begin{figure*}
\centering
\resizebox{0.9\textwidth}{!}
{\includegraphics{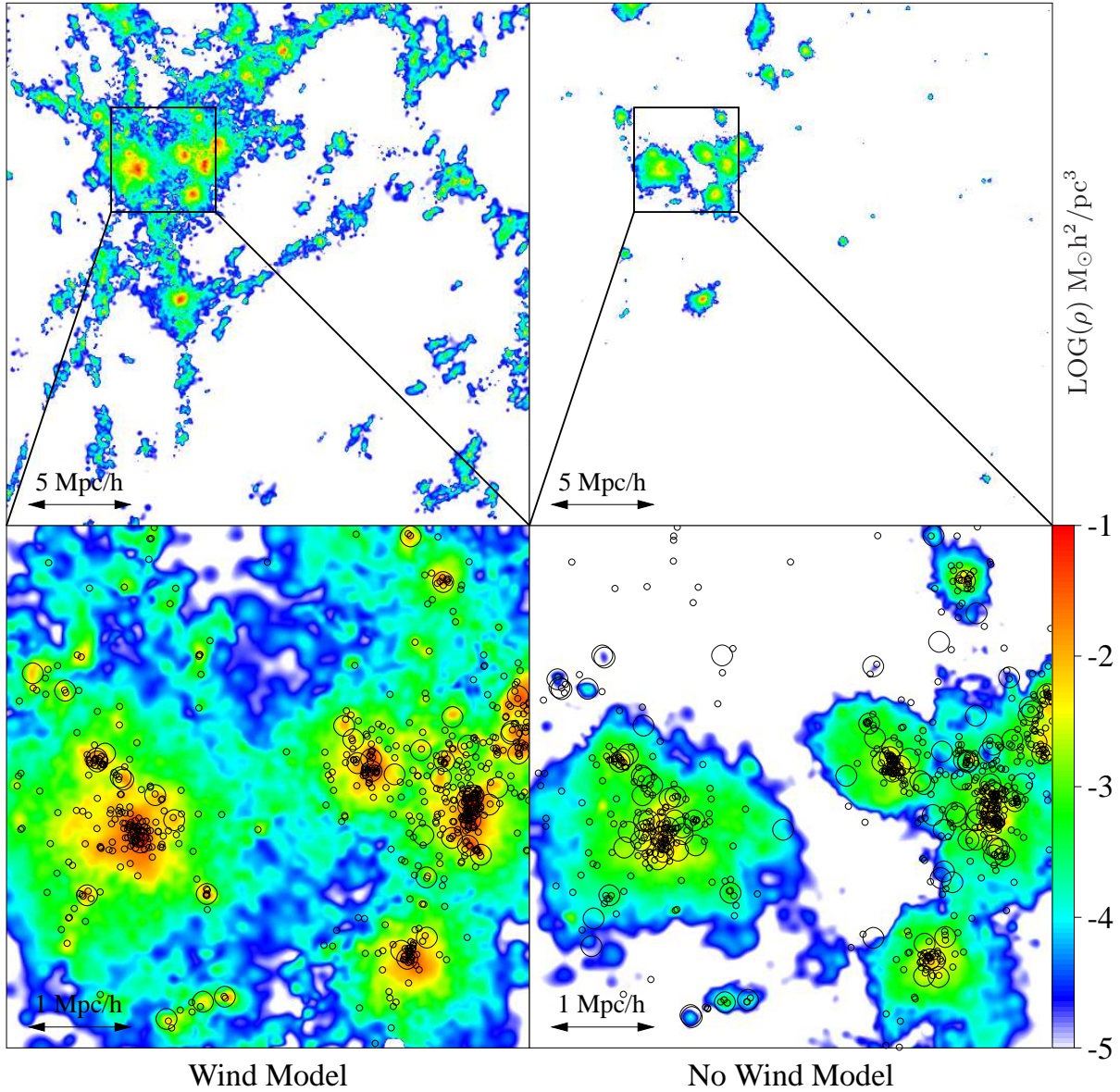}}
\caption{Surface density maps of the free metals~(i.e., metals not in galaxies)
in the Wind Model~(Left) and the No-Wind Model~(Right), at redshift $z=0.3$.
Top panels show a $25\hMpc\times25\hMpc\times5\hMpc$~(comoving) slice for each
model, with the density scale color coded as in the bottom-left color bar.
Bottom panels show $(5\hMpc)^3$ cubes zoomed into the densest region of the top
two panels. Large circles of $100\hKpc$ radius indicate the positions of
SKID groups more massive than $5.4\times10^{10}\Msol$, and small circles of
$30\hKpc$ radius mark those less than $5.4\times10^{10}\Msol$.} 
\label{fig:metalmap} 
\end{figure*}
%
\begin{figure*}
\centering
\resizebox{1.0\textwidth}{!}{\includegraphics{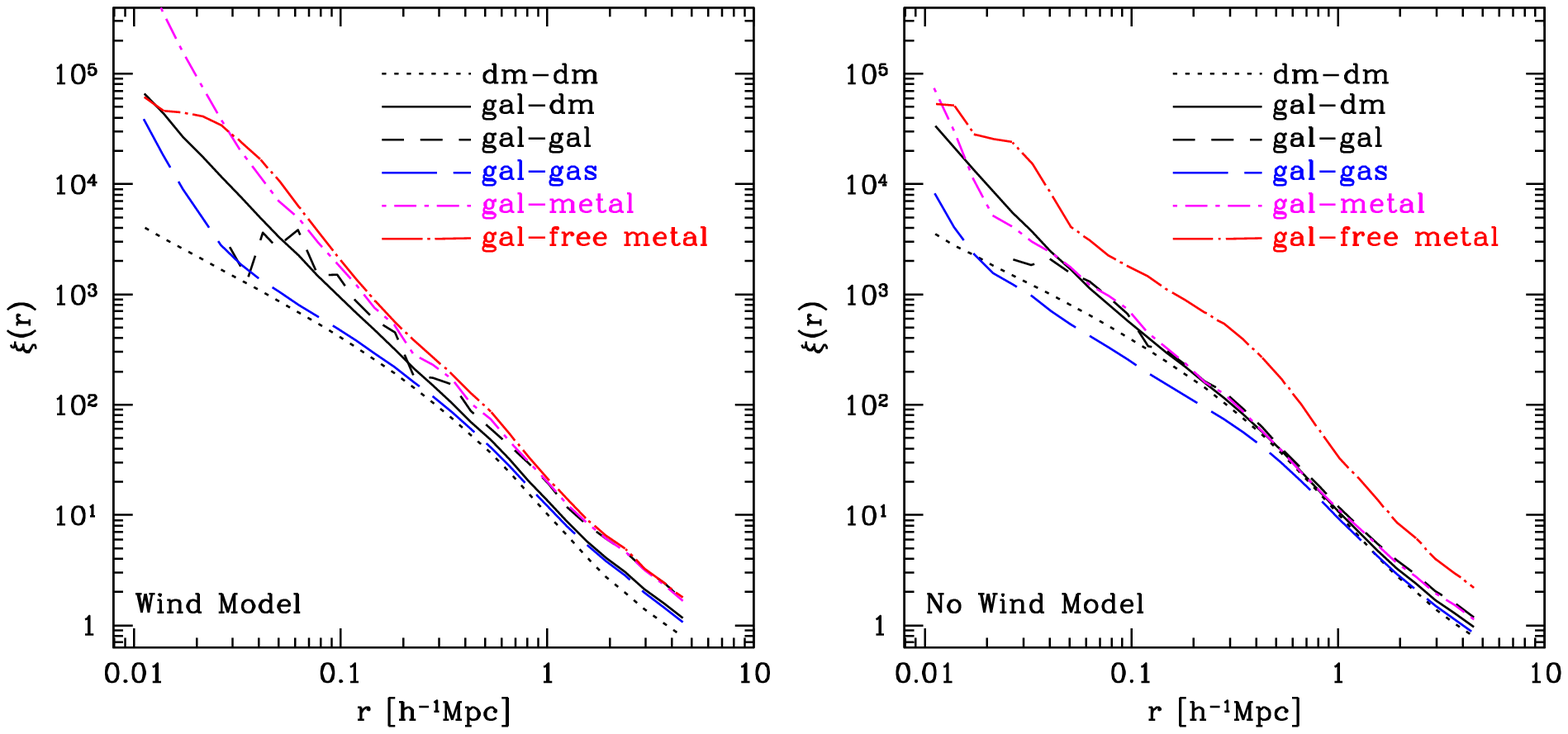}}
\caption{3-dimensional correlation functions in the Wind Model~(left) and
No-Wind Model~(right). Black dotted, solid, and dashed lines show the dark
matter autocorrelation, galaxy-dark matter cross-correlation, and galaxy
autocorrelation functions, respectively. Blue dashed lines show the galaxy-gas
cross-correlation. Magenta and red dot-dashed lines show the cross-correlation
of galaxies with all metals~(short dashes) and free metals~(long dashes),
respectively.}
\label{fig:xi} 
\end{figure*}
%

Four heavy elements~(C, O, Si and Fe) are tracked in the simulation. In this
exploratory work, we simply choose the total gas-phase metallicity as the
tracer of cosmic dust, because the amount of dust should be proportional to
the metal available if we assume the relative proportions of refractory
elements depend weakly on enrichment. We define ``free metals'' to be those in
gas particles that are not associated with {\it any} {\rm SKID} group (i.e.,
with no mass cut applied); recall that the SKID density threshold
is $\rho/\bar{\rho}_{\rm baryon} > 1000$, low enough to include gas in
extended regions of the interstellar medium.
We generally include only these ``free metals''
when computing the intergalactic dust extinction for comparison to MSFR, on
the assumption that quasars behind the optically thick disks of galaxies will
not make it into the SDSS sample because of obscuration.  The disks of bright
galaxies cover a small fraction of the sky in any case. However, for the
No-Wind simulation we also consider a ``hybrid dust'' calculation in which we
include metals in galaxies below $5.4\times10^{10} \Msol$, allowing the
possibility of the MSFR signal arising in low luminosity galaxies.  

Table~\ref{tab:tab1} lists the total mass of all gas-phase metals 
(interstellar plus intergalactic) and all free metals (intergalactic only) in
the two simulations. 
The total gas-phase metal mass is nearly identical in the
two cases.  This agreement is somewhat coincidental, as the No-Wind simulation 
forms more stars but leaves a larger fraction of its metals locked up in stars
and stellar remnants.  The mass of {\it free} metals in the Wind simulation
is 12
times higher, as much of the enriched gas is ejected from galaxies.  The top
panels of Figure~\ref{fig:metalmap} show the projected density distribution of
free metals from the two models in a $25\hMpc\times25\hMpc$ slice of $5\hMpc$
thickness; the lower panels show a $5\hMpc\times5\hMpc$ zoom on the densest
region of the slice. In these densest regions the free metal distributions of
the two models are fairly similar, tracing the galaxy distributions inside
groups and clusters.  Intergalactic metals in the No-Wind simulation presumably
come from a combination of tidal and ram pressure stripping, an enrichment
process first highlighted by~\cite{GO97}.  However, free metals in the No-Wind
simulation arise {\it only} in these dense group and cluster environments, where
stripping mechanisms can operate effectively.  The Wind simulation 
shows much more
widely distributed metals, in smaller halos and in the filaments that connect
them. Thus, the two simulations differ dramatically in the total amount of
free metals and in the spatial distribution of these metals. 

Large~(small) circles in the lower panels mark zones of radius
$100\hKpc$~($30\hKpc$) around SKID groups more~(less) massive than
$5.4\times10^{10}\Msol$, which we will refer to in our analysis below. The
spatial distributions of SKID groups are similar in the two simulations, though
masses are systematically lower in Wind simulation, 
converting some large circles
to small circles.
%
\section{Galaxy-Dust Correlations and Quasar Reddening} \label{sec:res}

Figure~\ref{fig:xi} shows 3-dimensional auto- and cross-correlation functions
from the two simulations. 
For the cross-correlation functions, we compute the 
(dark matter, gas, metal, or free metal) mass of neighboring
particles in spherical shells of successive radii, ranging from $10\hKpc$ to
$5\hMpc$ with logarithmic intervals, average over all galaxies above
our $5.4\times 10^{10} M_\odot$ threshold, and normalize by the
mass expected in a randomly located shell of equal volume.
The two dark
matter autocorrelation functions are nearly identical, as expected since the
simulations have the same initial conditions and the gravitational impact of
wind feedback is tiny. The galaxy autocorrelation and galaxy-dark matter
cross-correlation  are slightly higher in the Wind Model~(compare to the dark
matter autocorrelation at large scales) because in this simulation our
$5.4\times10^{10}\Msol$ mass threshold picks out rarer objects with a stronger
clustering bias. 

Turning to collisional components, the galaxy-gas correlation in the No-Wind
Model tracks the dark matter autocorrelation on scales $>1\hMpc$, then falls
below it at $r<0.5\hMpc$ until rising steeply inside 20$\hKpc$ because of
dissipational condensation into disks.  The deficit at intermediate scales
probably reflects the removal of gas from the inner regions of halos by this
condensation, and its subsequent conversion to stars.  The galaxy-gas
correlation in the Wind Model tracks the dark matter autocorrelation until the
sharp upturn of the former inside 30$\hKpc$.  The absence of an
intermediate-scale deficit could be an effect of gas redistribution by winds,
or it could reflect the lower amount of stellar conversion in this simulation.
The galaxy-metal correlations trace the galaxy-galaxy correlations beyond
0.5$\hMpc$ in both simulations, which is unsurprising since most of the metals
reside in galaxies. However, in the No-Wind Model the cross-correlation of
galaxies with {\it free} metals is much higher amplitude, by roughly a factor
of two at all scales, because the free metals in this model arise only in
group and cluster regions, which are highly biased relative to galaxies and
dark matter.

To facilitate comparison with observations, we concentrate on the direct
observable in the MSFR paper, the average reddening profile around galaxies,
$\extbar$, inferred from the change in quasar colors as a function of projected
separation. Note that we adopt comoving transverse distance $r_p$ rather than
the angular/physical distance in the MSFR paper. To estimate $\extbar$, we
first calculate the projected correlation function $\wp$ between galaxies and
free metals in the two models by directly summing metals through the simulation
cube in annuli around galaxies, using $x$, $y$, and $z$ projections. We then
convert $\wp$ to an average free-metal density profile, which we convert to an
excess dust surface density profile $\md$ by multiplying  by a dust-to-metal
ratio $R_{d/m}$. We convert $\md$ to a rest-frame reddening profile $\extbar$
assuming SMC-type dust as justified in MSFR, specifically 
\begin{equation}
\extbar \simeq 1.52 \langle_{\rm }E(B-V)_{\rm }\rangle_g =
 \frac{3.8}{{\rm ln}(10)}\frac{{\rm K}_{\rm ext}(\lambda_{\rm V})}{R_{\rm
V}}\md ~,
\end{equation}
where the factor $1.52$ comes from the conversion between two colors adopting
an extinction law $A_\lambda\propto\lambda^{-1.2}$ and K$_{\rm
ext}(\lambda_{\rm V})\simeq1.54\times10^{4}\,$cm$^2$g$^{-1}$ is the absorption
cross section per mass of SMC dust in V-band.  We choose the value of $R_{d/m}$
to match the MSFR data point at $r_p = 1 \hMpc$. While our standard calculation
simply assigns all the metals associated with a particle to the annulus in
which it resides, we have checked that smoothing the metals over the SPH kernel
before projecting yields indistinguishable reddening profiles.
%
%
\begin{figure*} 
\centering 
\resizebox{1.0\textwidth}{!}{\includegraphics{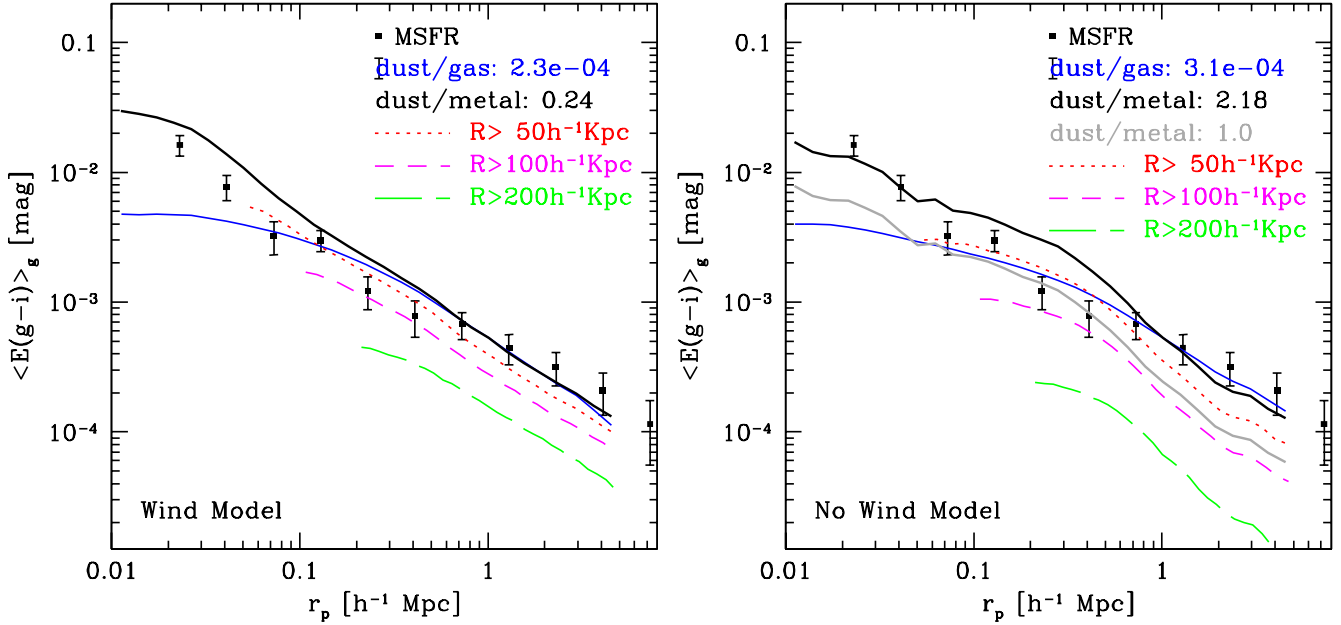}}
\caption{Average quasar reddening $\extbar$ as a function of comoving
transverse distance $r_p$ to a foreground galaxy in the Wind Model~(Left) and
the No-Wind Model~(Right). In each panel, solid squares with error bars are
the observational results from~MSFR.  The blue (black) solid line is the
predicted reddening profile assuming that dust traces gas (free metals), both
normalized by matching the MSFR data point at $1\hMpc$, with the dust-to-gas
or dust-to-metal ratio indicated in the legend. Red dotted, magenta dashed
and green long-dashed lines show the contribution to the reddening profile by
metals respectively $50\hKpc$, $100\hKpc$, and $200\hKpc$ away from galaxies
with baryonic mass $M>5.4\times 10^{10}M_\odot$. Reproducing the MSFR
normalization in the No-Wind Model with SMC duct requires an unphysical
dust-to-metal ratio of $2.18$. The gray solid line in the right panel is the
reddening profile for the No-Wind Model with SMC dust and a dust-to-metal
ratio of unity.}
\label{fig:dtom} 
\end{figure*}
%

Figure~\ref{fig:dtom} shows the central result of this paper, predicted
reddening profiles~(solid black curves) for the Wind Model~(left) and No-Wind
Model~(right), in comparison to the MSFR data points~(from their Figure~6). By
construction, the model curves go through the MSFR data point at $r_p =
1\hMpc$. For the Wind Model, this normalization requires $R_{d/m}=0.24$, i.e.,
$24\%$ of the free metal mass is in the form of dust. In the Milky Way
interstellar medium, roughly 50\% of the metal mass is in
dust \citep{Draine09}, so $R_{d/m}=0.24$ is physically plausible, but it
requires that a large fraction of the dust formed in the ISM survive the
journey to and sojourn in intergalactic space. Given the normalization at
$1\hMpc$, the model reproduces the shape of the MSFR profile fairly well from
$r_p=20\hKpc$ out to $r_p=5\hMpc$, beyond which box-size effects can
artificially depress the model predictions. MSFR show empirically that the
galaxy-dust correlation function approximately follows the galaxy-mass
correlation function, which is what we find theoretically for the galaxy-free
metal correlation function~(Figure~\ref{fig:xi}), so the approximate agreement
of shape is unsurprising.

For the No-Wind Model, the free metals would be insufficient to explain MSFR
reddening even if all of them were depleted onto dust~(gray solid line on the
right panel). Matching the MSFR normalization requires $R_{d/m}=2.18$, i.e.,
$218\%$ of the metal mass in dust~(black solid line on the right panel). This
is physically impossible, of course, but a change in the dust grain size
distribution from the SMC dust assumed here could possibly achieve more
efficient reddening for a given amount of metal mass. In detail, neither
model perfectly reproduces the shape of the MSFR data at $r_p<1\hMpc$, where
the ``1-halo'' regime of clustering begins to dominate over the ``2-halo''
regime~(see, e.g., Peacock \& Smith 2000; Berlind \& Weinberg 2002). The
agreement is worse for the No-Wind Model, where the halos hosting most of the
intergalactic dust are typically larger~(see Figure~\ref{fig:metalmap}).
However, when we analyze the three box projections individually, we find
$\sim 30\%$ variations from one to another, so we do not attribute much
weight to these discrepancies at present; we are also comparing predictions
at fixed redshift to inferences from angular correlations over a range of
redshift. Blue curves in Figure~\ref{fig:dtom} show the predicted reddening
profiles in the two models if we assume that dust traces intergalactic gas
rather than intergalactic metals. The MSFR normalization can be reproduced
for dust-to-gas mass ratios of $2.3\times10^{-4}$~(Wind Model) or
$3.1\times10^{-4}$~(No-Wind Model), comparable to those inferred from
observations of dwarf galaxies~\citep{LF98} and nearby
galaxies~\citep{Issa90}.  However, the dust-traces-gas reddening profiles
flatten inside $r_p\approx0.1\hKpc$, in contrast to the dust-traces-metal
profiles and the MSFR data points. In principle, observations like those of
MSFR can test hypotheses about the enrichment profiles of gas in galactic
halos and the survival of dust as a function of distance from the source
galaxy.

Dotted, short-dashed, and long-dashed lines in Figure~\ref{fig:dtom} show the
reddening profiles measured from the simulations if we eliminate sightlines
that pass within 50, 100, or 200$\hKpc$, respectively, of a galaxy above our
$5.4\times10^{10}\Msol$ threshold, keeping the dust-to-metal ratio the same.
(Circles in Figure~\ref{fig:metalmap} show the 100$\hKpc$ exclusion zones for
comparison.) In the Wind Model, dust outside these three exclusion zones
contributes $75\%$, $50\%$, and $30\%$ of the overall reddening signal at
large scales. In the No-Wind Model, the intergalactic dust is more tightly
clumped around the massive galaxies, and the reddening signal drops more
rapidly with radial exclusion, to $68\%$, $34\%$, and $12\%$, respectively.
In addition to characterizing the radial scale of intergalactic dust in our
simulations, these curves provide testable predictions of our models, which
can be implemented observationally by eliminating quasars with small
projected separations from foreground galaxies before computing the mean
reddening profile.

\begin{figure*} 
\centering
\resizebox{1.0\textwidth}{!}
{\includegraphics{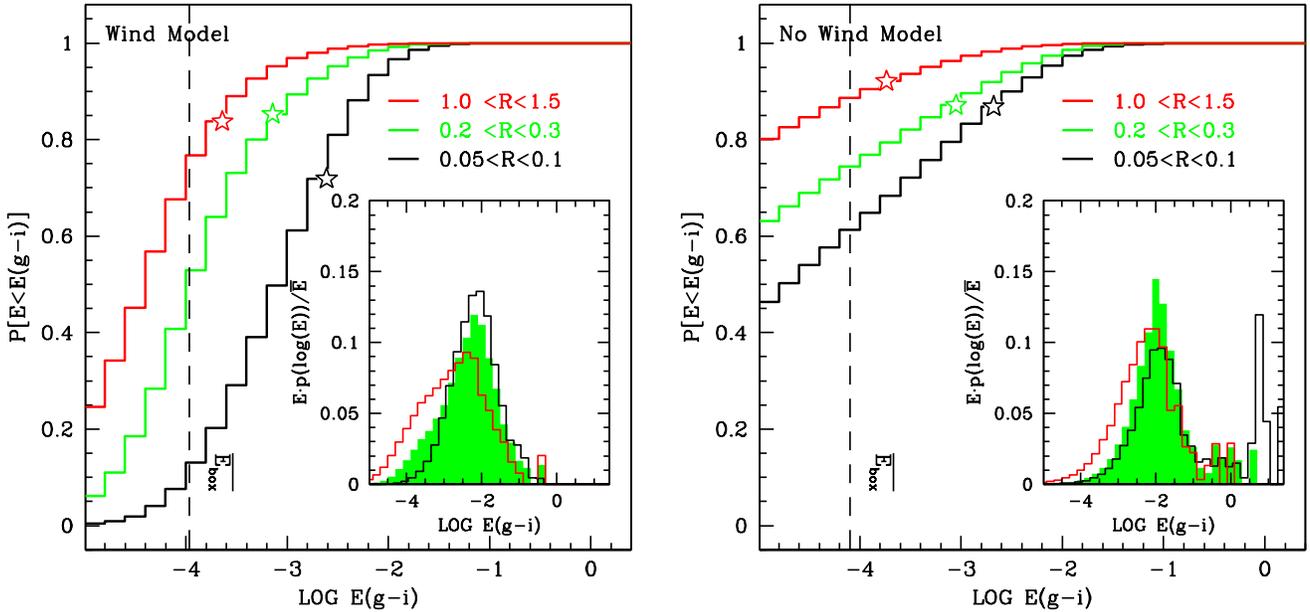}}
\caption{Cumulative distribution functions~(CDFs) of reddening for three radial
bins in the Wind Model~(Left) and No-Wind Model~(Right).  In each panel, the
black histogram shows the reddening CDF of quasars with projected separation
$0.05\hMpc < R < 0.1\hMpc$ from a foreground galaxy, while green and red
histograms show corresponding CDFs for separations $0.2\hMpc < R < 0.3\hMpc$ and
$1.0\hMpc < R < 1.5\hMpc$, respectively. The black dashed line indicates the
average amount of reddening through each simulation box, and the star on top of
each histogram indicates the mean reddening value for the radial bin. Inset
panels show histograms of $E\cdot_{\rm }p(\log_{\rm }E)/{\bar E}$,  the
fractional contribution from each $\log_{\rm }E(g-i)$ bin to the average amount
of reddening marked by stars in the CDF. The histogram of each radial bin is 
indicated by the same color as in the main panel.}
\label{fig:reddendist} 
\end{figure*}

Given the distinctive spatial patterns of the dust distributions in
Figure~\ref{fig:metalmap}, we expect the cumulative distribution
functions~(CDFs) of reddening to be quite different in the two models, with
more of the total extinction in the Wind Model arising at very low $E(g-i)$
and more sightlines in the No-Wind Model being truly dust free.
Figure~\ref{fig:reddendist} confirms this expectation.  We calculate each
reddening CDF by randomly picking sightlines within a given radial range
around any galaxy (not necessarily the closest galaxy) on a projected
reddening map generated by the {\it vista}\footnote{{\it vista} creates a fits
image whose pixel values are smoothed quantities calculated by projecting the
particles onto the grid using the SPH spline softening kernel.} command in
{\rm TIPSY}\footnote{http://hpcc.astro.washington.edu/tools/tipsy/tipsy.html}.
The black histogram in each panel shows the reddening CDF in the projected
separation bin at $50\hKpc<R<100\hKpc$, while the green and red histograms
show $200\hKpc < R < 300\hKpc$ and $1.0\hMpc<R<1.5\hMpc$, respectively. The
star on each histogram indicates the mean $E(g-i)$ value in the corresponding
radial bin, and the black dashed line shows the mean amount of reddening in
each simulation box.  These mean extinction values are proportional to the
total amount of dust in the simulation box, and thus to the comoving box
length of simulation.   In the Wind Model, the median reddening is $\approx
10^{-3}$ mag in the innermost radial bin, dropping to $10^{-4}$ mag and
$10^{-4.4}$ mag in the next two bins.  In the No-Wind Model, on the other
hand, most sightlines are nearly dust free, with median reddening below
$10^{-5}$ mag in all three radial bins.  Only 3\% of sightlines in the
innermost bin for the Wind Model have reddening greater than 0.01 mag, and
only 3.5\% in the No-Wind Model.  Unfortunately, because nearly all sightlines
have a reddening level that is small compared to the intrinsic dispersion of
quasar colors, the large differences in the CDFs of the two models are likely
to be unobservable in practice.
%

The inset panels in Figure~\ref{fig:reddendist} show histograms of
$E\cdot_{\rm }p(\log_{\rm }E)/{\bar E}$, the fractional contribution to the
average $E(g-i)$ from each bin shown in the CDF.  In the Wind Model,
contributions to the mean reddening are broadly peaked in the range $E(g-i) =
0.001-0.03$ mag, shifting towards lower values for the largest ($1-1.5\hMpc$)
radial separation bin.  Contributions for the No-Wind Model also peak at 0.01
mag, though they continue up to strongly reddened sightlines with
$E(g-i)=0.1-0.6$ mag.  However, even if these noticeably reddened quasars were
eliminated from the sample (deliberately or by selection bias), the mean
reddening profile would barely change.

\begin{figure*} 
\centering
\resizebox{1.0\textwidth}{!}
{\includegraphics{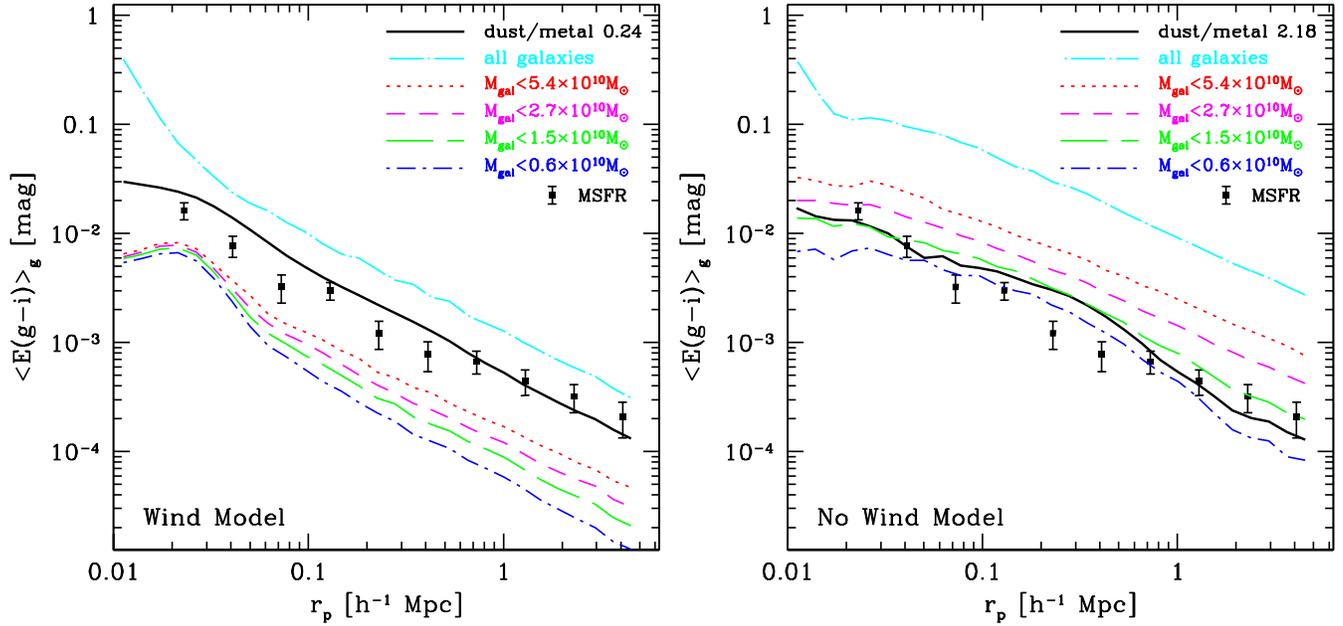}}
\caption{Reddening caused by dust in low mass galaxies in the Wind
Model~(Left) and No-Wind Model~(Right). In each panel, solid squares are
from~MSFR, and solid black lines show the reddening produced by intergalactic
gas assuming a constant dust-to-metal ratio on all scales as in
Fig.~\ref{fig:dtom}.  Blue dot-dashed, green long-dashed, magenta dashed,
and red dotted, lines indicate the additional reddening when including dust
contained by galaxies with masses less than 0.6, 1.5, 2.7, and $5.4\times
10^{10}\Msol$, respectively.  The sum of one of these curves with the black
curve shows the reddening predicted if the MSFR sample included quasars
behind galaxies lower than the corresponding mass threshold, assuming the
same dust-to-metal mass ratio both inside and outside of galaxies.  Cyan
dot-dashed lines (the highest in each panel) show the reddening from dust in
{\it all} galaxies.  }
\label{fig:dwarfext} 
\end{figure*}

Figure~\ref{fig:dwarfext} compares the correlated reddening signal from
intergalactic dust (tracing free metals) to the signal that would be
contributed by including dust from galaxies below a succession of baryonic
mass thresholds.  We assume the same dust-to-metal mass ratio inside and
outside of galaxies.  Beyond $r_p \approx 30\hKpc$, the shapes of the
correlated reddening profiles are all similar, tracking the shapes of the
galaxy-galaxy and galaxy-mass correlation functions.  When we consider dust in
all galaxies (cyan dot-dashed line), the normalization for the No-Wind Model
is ten times higher than that of the Wind Model, but this difference just
reflects the higher dust-to-metal ratio assumed to match the free-metal
prediction to the MSFR data.  If we eliminate galaxies with baryonic mass $M >
5.4\times 10^{10} M_\odot$, the reddening signal drops by a factor of six in
the Wind Model but only a factor of three in the No-Wind Model.  This
difference reflects the greater efficiency of winds in ejecting metal-enriched
gas from lower mass galaxies.  If we consider only galaxies below $5.4\times
10^{10} M_\odot$, the galactic dust is considerably less than the
intergalactic dust in the Wind Model, but considerably more in the No-Wind
Model.  While the No-Wind Model has several times more mass in galaxies,
and a higher space density of galaxies above a given mass threshold,
the amount of metal mass (and hence dust) in the two simulations
is similar (Table~\ref{tab:tab1}).  The model differences in 
Figure~\ref{fig:dwarfext} arise from the different distributions
of metals inside and outside galaxies.

\begin{figure*} 
\centering
\resizebox{1.0\textwidth}{!}
{\includegraphics{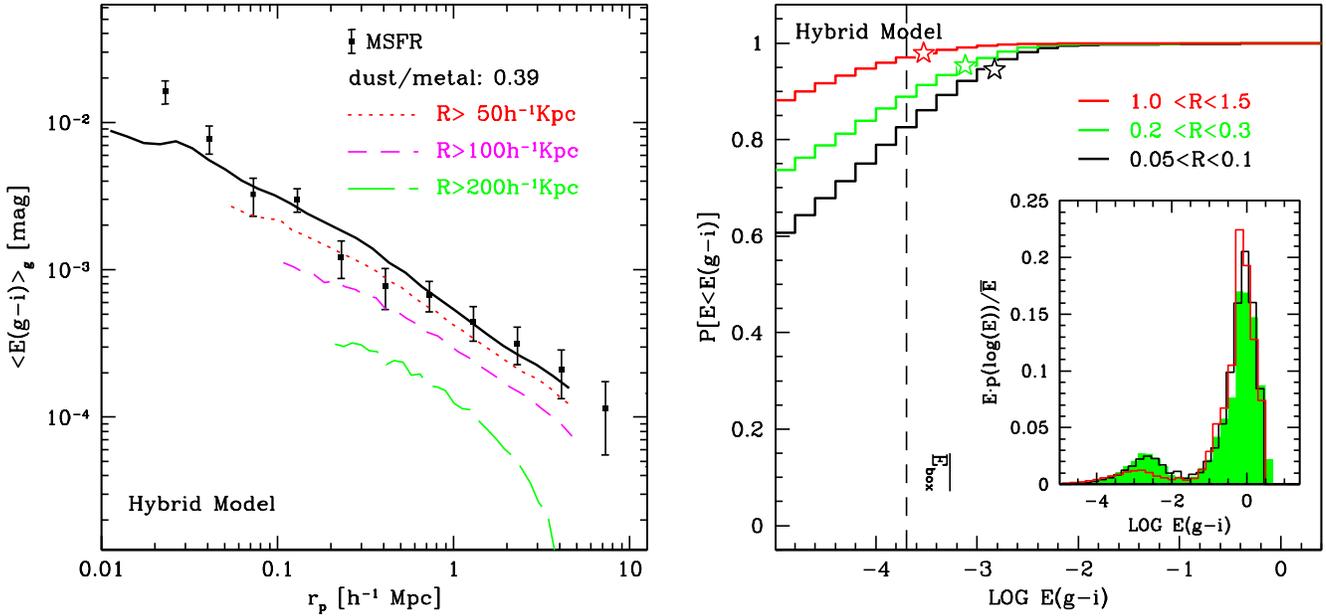}}
\caption{An alternative ``hybrid'' dust model including dust within low mass
galaxies ($M< 5.4 \times 10^{10} M_\odot$) in the No-Wind simulation. {\it Left
panel:} Comparison between extinction from hybrid dust (black solid line) and
MSFR results~(solid points). Red dotted, magenta dashed and green long-dashed
lines indicate the extinction contribution from dust $50\hKpc$, $100\hKpc$ and
$200\hKpc$ away from high mass ($M>5.4 \times 10^{10} M_\odot$) galaxies,
respectively.  {\it Right panel:} Cumulative distribution functions of
reddening for the hybrid dust model. Black, green and red histograms show
reddening CDFs of quasars in three bins of radial separation (from high mass
galaxies), respectively $0.05-0.1$, $0.1-0.2$, and $1.0-1.5\hMpc$.  The
vertical dashed line indicates the average amount of reddening through the box,
while the stars mark the mean extinction value for each radial bin. Inset panel
shows the $E\cdot_{\rm }p(\log_{\rm }E)/{\bar E}$ for each radial bin, with the
same format as Figure~\ref{fig:reddendist}, but larger $y$-axis range.}
\label{fig:fmet2} 
\end{figure*}

\begin{figure} 
\centering 
\resizebox{0.85\columnwidth}{!}{\includegraphics{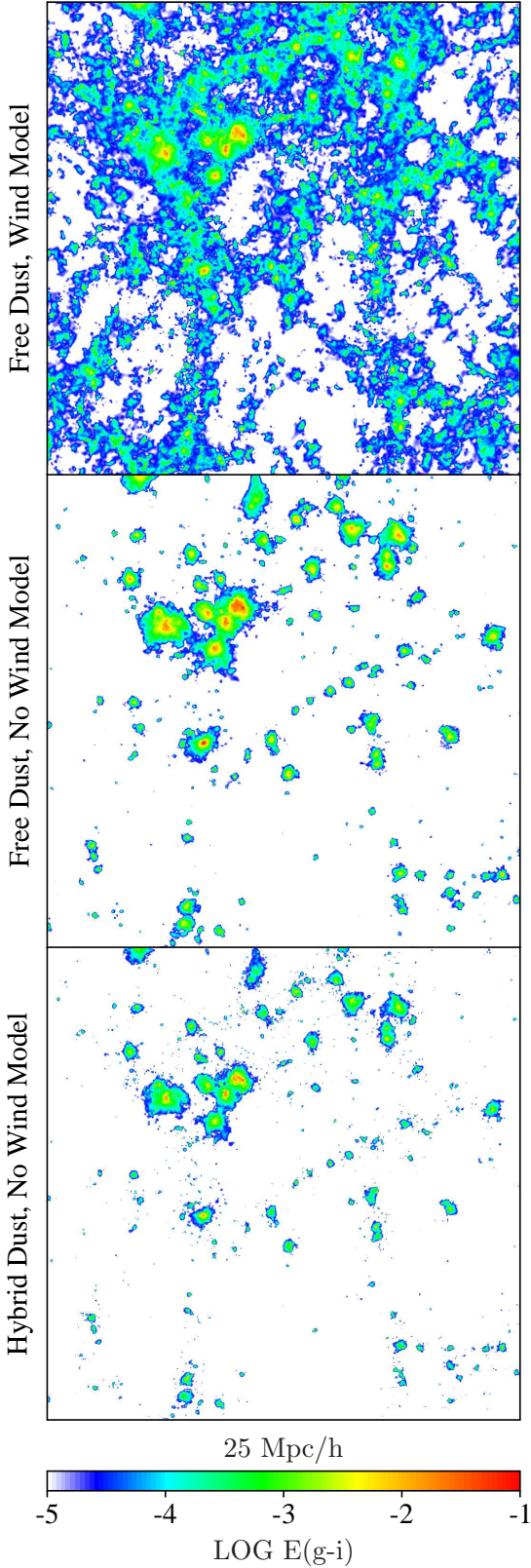}}
\caption{Comparison of reddening maps in the three dust models, each normalized
to the MSFR result at 1$\hMpc$.  The regions shown are a quadrant of the
simulation cube, $25\hMpc\times25\hMpc\times50\hMpc$.  Red/blue regions
indicate heavily/lightly reddened fields through the simulation, with the color
scale running from $E(g-i)=10^{-5}$ up to $0.1$, logarithmically.}
\label{fig:extmap} 
\end{figure}

With intergalactic dust alone, the No-Wind Model requires an unphysical ratio
2.18 of dust mass to metal mass to reproduce the MSFR data.  However,
Figure~\ref{fig:dwarfext} shows that low mass galaxies contain much more
metal mass than the intergalactic medium in the No-Wind Model. We therefore
construct an alternative ``hybrid dust'' model from the No-Wind simulation,
in which we include both dust associated with free metals and dust associated
with low mass galaxies. The upper limit for the baryonic mass of those
galaxies is $\sim \rm{few} \times10^{10} M_\odot$, largely determined by the
total amount of metal (both free and inside faint galaxies) that is required
to achieve a reasonable dust-to-metal ratio. For instance, if we include
dust associated with galaxies with $M < 1.5\times10^{10}M_\odot$, the
``hybrid dust'' model would require $87\%$ of the metals in the form of dust.
The required dust-to-metal ratio drops to $60\%$ if we increase the upper
limit of low mass galaxies to $2.7\times10^{10}M_\odot$. To obtain a more
reasonable dust-to-metal ratio ($0.1<R_{d/m}<0.5$), we need an upper limit of
$\sim 5 \times 10^{10} M_\odot$. For the sake of uniformity of mass
thresholds in the paper, we set $M=5.4 \times 10^{10} M_\odot$ again to be
the maximum baryonic mass of ``faint'' galaxies.  Figure~\ref{fig:fmet2}
shows the reddening profile and reddening CDF for this hybrid model, and
Figure~\ref{fig:extmap} compares the extinction map of this model to those of
the Wind and No-Wind models.  Reproducing the normalization of the MSFR
measurements requires a physically acceptable dust-to-metal mass ratio of
0.39.  The effect of excluding zones around high mass galaxies is similar to
that in the Wind Model (compare the broken lines in the left panels of
Figure~\ref{fig:fmet2} and Figure~\ref{fig:dtom}), except for a precipitous
drop in the $R > 200\hKpc$ curve beyond $r_p=3\hMpc$.  The reddening CDF is
similar to that of the No-Wind Model (compare the right hand panels of
Figures~\ref{fig:reddendist} and~\ref{fig:fmet2}).  While the addition of low
mass galaxies reduces the number of truly metal-free sightlines, the factor
of 5.5 reduction in the dust-to-metal mass ratio compensates by shifting the
previously obscured sightlines to lower reddening values.  One can see this
behavior in the extinction map of Figure~\ref{fig:extmap}: the extended blobs
of free metals in groups and clusters shrink because of the lower dust
normalization (the outer contours lie at a fixed reddening threshold of
$10^{-4}$), and they are supplemented by a peppering of dots that show the
low mass galaxies tracing filamentary superclusters.  
%
%

From the inset panel of Figure~\ref{fig:fmet2}, one can see that the
reddening distribution of the hybrid model is bimodal, with a
``galactic'' peak at high reddening and an ``intergalactic'' peak
at low reddening.
While sightlines with $E(g-i)>0.1$ are rare,
they contribute a large fraction of the mean reddening.
These high reddening values might
change quasar colors or magnitudes enough to throw them out of
the quasar catalog used by MSFR for their measurement.  Moreover, the light of
the foreground galaxy itself might eliminate the background quasar from the
input catalog, either by changing its color or by adding a detectable extended
component that changes the object classification.  At $z=0.3$, a galaxy with
baryonic mass $5.4 \times 10^{10} M_\odot$ has apparent magnitude $g \approx
20.4$, assuming a stellar mass-to-light ratio $M/L_g = 3 M_\odot/L_\odot$,
while the limit of the quasar catalog used by MSFR is $g \approx 21$.  Thus,
at least the brighter galaxies in the hybrid model would likely have a
significant impact.  We will not address the detailed selection questions
here, but the clearest test of the hybrid model would be to search for
signatures of blended galaxy light (perhaps by stacking the images of the
selected quasars), as a function of separation from the bright foreground
galaxies used for the cross-correlation measurement.

MSFR argue that the dust in LMC-like dwarfs is insufficient to explain the
magnitude of their reddening signal.  We concur with this conclusion.
Reproducing the MSFR data in the hybrid model with a physical dust-to-metal
mass ratio requires including galaxies up to several times $10^{10} M_\odot$,
far larger than the $\sim 3\times10^9 M_\odot$ baryonic mass of the
LMC~\citep{Marel02}.  Furthermore, the No-Wind simulation predicts a galaxy
baryonic mass function that is inconsistent with observations, with an
excessive global fraction of baryons converted to
stars~\citep{Oppenheimer10}\footnote{The Wind simulation predictions are
reasonably consistent with the observed mass function for galaxies with
$L<L_*$, though it still predicts excessive galaxy masses above $L_*$.}.  We
do not consider the hybrid dust model to be nearly as plausible an explanation
of the MSFR results as the Wind Model; we present it as a foil to illustrate
what would be required to explain MSFR's findings with dust in low mass
galaxies.  For the Wind Model, the metals in low mass galaxies contribute much
less reddening than the intergalactic metals (Figure~\ref{fig:dwarfext}).  
%
%

%
\section{Discussion} \label{sec:dis}
%
For the Wind Model to succeed, we require that the dust-to-metal mass ratio in
the IGM be comparable to that in the ISM, allowing only $\sim 50\%$ of the ISM
dust to be destroyed during its expulsion from galaxies and subsequent
residence in the IGM.  The validity of this assumption is by no means obvious,
as the destruction timescales for $0.01\,\mu$m dust grains by thermal
sputtering are $\sim 10^{7.5}(n_H/10^{-3}\cm^{-3})^{-1}$ years at $T=10^6\K$
(\citealt{DS79}, Figure 7), while wind particles in the simulation typically
remain in the IGM for $\sim 10^9$ years before reaccreting onto galaxies
(\citealt{Oppenheimer10}, Figure 2).  However, the sputtering rates decline
rapidly towards lower temperatures (e.g., a factor of 300 lower at
$T=10^5\K$), and with the wind implementation used in this simulation most
ejected gas never rises above a few $\times 10^4\K$.  UV or X-ray background
photons are another possible destruction mechanism for IGM dust, but the
intergalactic radiation field is much lower intensity than the radiation field
dust grains already encounter in galactic star-forming regions.

A detailed consideration of dust survival in the IGM is beyond the scope of
this initial investigation, but the MSFR results clearly raise it as an
important subject for further study.  The combination of their measurements
with our models gives a fairly clear idea of what is required: survival of a
substantial fraction of ejected dust, and an extinction curve that has roughly
the color dependence of ISM dust.  The temperature sensitivity of thermal
sputtering could lead to preferential destruction of ejected dust in the
higher mass halos that host a shock heated gas halo (see
\citealt{Birnboim03,Keres05,Dekel06,Keres09a}).  In the Wind Model, most wind
particles in halos with $M < 10^{13} M_\odot$ have $T < 10^5\K$, but about 2/3
of the wind particles in halos with $M>10^{13} M_\odot$ have $T > 3\times
10^6\K$.  If sputtering does destroy intergalactic dust at these temperatures,
it could produce distinctive drops in the galaxy-reddening correlation when it
is evaluated for massive galaxies or for galaxies in dense environments.  The
recent study of \cite{McGee10}, which examines the correlation of background
quasar colors with projected separation from galaxy groups of varying
richness, provides some hint of such an effect, but their innermost point is
at $r = 1\hMpc$, close to the virial radius of typical group mass halos.
Moreover, the \cite{Chelouche07} measurements provide direct evidence for dust
survival in the cluster IGM.  \cite{Draine09} argues that ISM dust must form
largely {\it in situ}, from the depletion of gas phase metals onto seed grains
from supernovae and ejected stellar envelopes, and that the dust abundance is
determined by an equilibrium between growth and destruction mechanisms.
Growth rates would be much slower in the low density IGM, but 2-body
destruction processes would have the same density scaling, so a similar
equilibrium abundance could arise.  
%
%

Intergalactic dust could have an important impact on supernova cosmology
studies, dimming and reddening supernovae increasingly with redshift.  Given
perfect knowledge of supernova intrinsic colors and the shape of the extinction
curve, intergalactic extinction would be corrected automatically along with
extinction in the supernova host galaxy.  However, the intergalactic and host
galaxy dust components could have different extinction curves, complicating the
analyses.  Furthermore, the presence of intergalactic dust means that no
high-redshift supernovae have zero extinction, contrary to the assumption
usually made in global models of the supernova population.  MSFR give a rough
estimate of the average extinction implied by their results, $\langle A_V
\rangle = 0.03$ mag at $z=0.5$, but the inference of a mean extinction from the
{\it correlated} reddening signal is 
(as MSFR emphasize) sensitive to uncertain choices of radial
profile cutoffs and luminosity extrapolations.
(See \citealt{Menard09b} for further discussion.)

Our simulation provides an explicit physical model of the IGM metal
distribution, and once the dust-to-metal mass ratio is set by matching the
MSFR data, it is straightforward to compute the mean extinction by
intergalactic dust.  In the Wind Model, the comoving dust density 
$\rho_{dust}=6.424\times10^{-35}$g\,cm$^{-3}$. Assuming constant comoving dust
density and extinction law $A_\lambda\propto\lambda^{-1.2}$, we can compute
mean extinction to any given redshift $z$ as
\begin{equation}
    \langle A_V \rangle = \frac{2.5\,K_{\rm ext}\,c}{{\rm
    ln}(10)}
    \int_0^z{(1+z)^{3.2}H(z)^{-1}\rho_{\rm dust}{\rm d}z},
\end{equation}
where $H(z)$ is the Hubble constant at redshift $z$, respectively. The
$(1+z)^{3.2}$ term arises from the combination of cosmic areal expansion and
rest-frame to observed V-band extinction conversion. This yields a predicted
mean extinction $\langle A_V \rangle = 0.0133$ mag and a
sightline-to-sightline variance of $\Delta^2 = (0.0272\,{\rm mag})^2$ for
supernovae at $z=0.5$. More accurate estimates and an extrapolation to
higher redshifts will require the combined analysis of many simulation
outputs to track evolution.  Our estimated extinction is small compared to
the statistical and systematic errors of current surveys, but it is
comparable to the ambitious goals set for next-generation surveys of several
thousand high-redshift supernovae.  At higher redshifts, furthermore, the
(stronger) near-UV dust extinction gets redshifted into the observed-frame
optical bands.  We have assumed SMC dust for this calculation; since we
normalize to MSFR's $E(g-i)$ measurements, the important assumption is the
value of $A_V/E(g-i) = 4.7$.
%
\section{Summary} \label{sec:conc}

Compared to the element-by-element and sightline-by-sightline measurements of
gas phase metals in quasar spectra, the correlation of galaxy separation and
quasar color used by MSFR is a relatively blunt tool, requiring tens of
thousands of background sources to yield a statistical detection of an
intrinsically weak effect, and providing only a radial dependence and absolute
normalization in several color bands.  Nonetheless, this tool provides unique
insight into an aspect of IGM enrichment that is virtually impossible to study
by other means.  We have investigated the ability of SPH cosmological
simulations with and without galactic winds to explain MSFR's remarkable
results.

The Wind Model, based on an implementation of galactic winds that has proven
empirically successful in other contexts and is motivated by theoretical models
of momentum-driven outflows, proves quite successful at reproducing the MSFR
measurements, provided that about 25\% of the metal mass in the IGM is in the
form of SMC-like dust.  This dust-to-metal mass ratio is roughly half that in
the Milky Way ISM, so this normalization is reasonable provided dust survives
its ejection from galaxies and its sojourn in the IGM.  
Accounting for dust depletion would lower the predicted level of
intergalactic {\it gas} absorption in quasar spectra by $\sim 25\%$,
at least for refractory elements such as carbon and silicon.
In our Wind simulation,
about 1/3 of the metals are intergalactic, and 2/3 are in galaxies (not
counting the metals locked up in stars), with 60\% of the latter in galaxies of
baryonic mass $M>5.4\times 10^{10} M_\odot$.  If we exclude sightlines that
pass within 50, 100, and $200\hKpc$ of galaxies above this mass threshold, then
the large scale correlated reddening signal drops to 75\%, 50\%, and 30\% of
the original signal, demonstrating the large scale of the distributed metals
and providing an observationally testable diagnostic of the model
(Figure~\ref{fig:dtom}).  The cumulative distribution function (CDF) of
reddening in each radial separation bin is a prediction that is testable in
principle, but not in practice because the overall extinction is always
dominated by sightlines with $E(g-i)< 0.03$ mag, small compared to the
intrinsic dispersion of quasar colors.

The No-Wind simulation has just 3\% of its metals in the intergalactic medium,
almost all of it in the dense halos of groups and clusters where tidal
interactions and ram pressure can strip enriched gas from member galaxies.  The
3-dimensional cross-correlation function of galaxies with free metals is highly
biased, but because the free metal fraction is so low to begin with,
reproducing the MSFR data requires an unphysical dust-to-metal mass ratio of
2.18 in the IGM.  This model could conceivably be made consistent with the data
if the grain size distribution were altered to give more reddening for a given
dust mass, though the consistency of the MSFR band-dependent measurements with
``standard'' interstellar dust limits any such effect.  Compared to the Wind
Model, radial exclusion around high mass galaxies has a stronger effect on
large scale correlations, and the CDF shifts to somewhat higher values, though
still too low to allow measurement of the CDF in the presence of quasar
color variations.
 
We have also considered a ``hybrid'' model that uses the No-Wind simulation but
includes reddening in galaxies with baryonic mass $M < 5.4\times
10^{10}M_\odot$ (which dominates the intergalactic dust signal by a factor of
$2-4$) when computing the large scale galaxy-dust correlation.  This model can
reproduce the MSFR measurements with a physically acceptable dust-to-metal mass
ratio of 0.39. 
However, while sightlines with $E(g-i)>0.1$ are still rare in this
model, they contribute a significant
fraction of the mean reddening.  Moreover,
the galaxies presumed to produce
most of the reddening in this model might well have detectable effects on
quasar colors, visual morphologies, or spectra, and no such effects have been
reported.  Predictions for this model are also more sensitive to our mass and
spatial resolution, and to the failure of the No-Wind simulation to match the
low mass end of the observed galaxy baryonic mass function
\citep{Oppenheimer10}.  In the end, we consider both the No-Wind and hybrid
models to be illustrative demonstrations of how hard it is to reproduce the
MSFR data {\it without} widespread galactic outflows.

The large scale galaxy-dust correlation measured by MSFR traces the shape of
the galaxy-mass correlation function they infer from gravitational
magnification, which in turn traces the known shape of the galaxy-galaxy
correlation function.  This large scale shape is almost inevitable in any model
where the dust originates in galaxies, and we find it in all three of our
models despite the very different spatial distributions of the dust.  However,
on sub-Mpc scales --- the 1-halo regime --- the detailed form of the
correlation function has substantial diagnostic power for the sources of the
dust, the extent and radial profile of outflows, and the survival of dust in
these outflows, albeit in a combination that may be difficult to untangle.  In
this regime, our three models predict significantly different shapes, none of
them in full agreement with the data, and the shape of the galaxy-metal
correlations is quite different from the shape of the galaxy-gas correlations.
Given the fairly large ($\sim 50\%$) observational error bars and our
simplified modeling of MSFR's angular measurements in terms of projected
separations at an effective redshift, we have not attempted to use the detailed
small-scale shape as a diagnostic or to assess the implications of moderate
discrepancies with our theoretical predictions.  However, future imaging
surveys such as Pan-STARRS, the Dark Energy Survey, and LSST will provide much
larger samples of quasars and foreground galaxies, allowing higher precision
measurements in shells of photometric galaxy redshift.  Over the past decade,
galaxy-galaxy lensing has progressed from the first measurements of large scale
galaxy-matter angular correlations \citep{Fischer00} to detailed measurements
of projected mass correlations as a function of galaxy luminosity, color,
morphology, and environment \citep{McKay01,Hoekstra04,Sheldon04,Mandelbaum06}.
We anticipate similar progress in the studies of galaxy-dust correlations over
the coming decade, providing new insights into the origin, evolution,
and observational impact of solid-phase material
in the intergalactic medium.
%
\section*{Acknowledgements} We thank 
Charlie Conroy,
Bruce Draine,
Brice M\'enard, 
Ryan Scranton,
Masataka Fukugita, and
Gordon Richards
for helpful discussions.  
We thank Vimal Simha for investigating the temperature distribution
of wind particles as a function of halo mass (see \S\ref{sec:dis}).
This work was supported in part by NSF
Grant AST-0707985 and by an Ohio State University Fellowship to YZ.  
DW, MF, and DK gratefully acknowledge support from,
respectively, an AMIAS membership at the IAS, the FCAD Fellows Program
at the University of Massachusetts, and Harvard University.
%
%

%
\end{document}